\documentclass[english]{article}
\usepackage[utf8]{inputenc}
\usepackage[T1]{fontenc}
\usepackage{babel}
\usepackage{amsmath}
\usepackage{graphicx}
\usepackage{url}
\usepackage{xcolor}
\usepackage{booktabs}
\usepackage{caption}
\usepackage{subcaption}
\usepackage{listings}
\usepackage{tabularx}

\definecolor{codegreen}{rgb}{0,0.6,0}
\definecolor{codegray}{rgb}{0.5,0.5,0.5}
\definecolor{codepurple}{rgb}{0.58,0,0.82}
\definecolor{backcolour}{rgb}{0.95,0.95,0.92}

\lstdefinestyle{mystyle}{
    backgroundcolor=\color{backcolour},   
    captionpos=b,                    
    keepspaces=true,                 
    showspaces=false,                
    showstringspaces=false,
    showtabs=false,                  
    tabsize=2
}
\definecolor{airforceblue}{rgb}{0.36, 0.54, 0.66}
\definecolor{antiquebrass}{rgb}{0.8, 0.58, 0.46}
\definecolor{applegreen}{rgb}{0.55, 0.71, 0.0}
\definecolor{bazaar}{rgb}{0.6, 0.47, 0.48}
\definecolor{brickred}{rgb}{0.8, 0.25, 0.33}
\definecolor{cambridgeblue}{rgb}{0.64, 0.76, 0.68}
\definecolor{chestnut}{rgb}{0.8, 0.36, 0.36}
\definecolor{darksalmon}{rgb}{0.91, 0.59, 0.48}
\definecolor{darkchampagne}{rgb}{0.76, 0.7, 0.5}

\usepackage{fancyhdr}

\usepackage{lastpage}
\usepackage{setspace}
\newcommand{\rquest}[1]{%
\vspace{5mm}\noindent\parbox{\textwidth}{\onehalfspacing\textbf{\large #1}\vspace{4mm}}
~\par%
}
\begin{document}

\title{A large-scale study on \\research code quality and execution }

\author{Ana Trisovic\textsuperscript{1{*}}, 
Matthew K. Lau\textsuperscript{2},
Thomas Pasquier\textsuperscript{3},
Merc\`e Crosas\textsuperscript{1}}

\maketitle
\thispagestyle{fancy}

\noindent 1. Institute for Quantitative Social Science, Harvard University, Cambridge, MA, USA\\
2. CAS Key Laboratory of Forest Ecology and Management, Institute of Applied Ecology, Chinese Academy of Sciences, Shenyang, China\\
3. Department of Computer Science, University of British Columbia, Vancouver, BC, Canada\\
{*}corresponding author(s): Ana Trisovic (anatrisovic@g.harvard.edu)


\begin{abstract}
This article presents a study on the quality and execution of research code from publicly-available replication datasets at the Harvard Dataverse repository. Research code is typically created by a group of scientists and published together with academic papers to facilitate research transparency and reproducibility. For this study, we define ten questions to address aspects impacting research reproducibility and reuse. First, we retrieve and analyze more than 2000 replication datasets with over 9000 unique R files published from 2010 to 2020. Second, we execute the code in a clean runtime environment to assess its ease of reuse. Common coding errors were identified, and some of them were solved with automatic code cleaning to aid code execution. We find that 74\% of R files crashed in the initial execution, while 56\% crashed when code cleaning was applied, showing that many errors can be prevented with good coding practices. We also analyze the replication datasets from journals' collections and discuss the impact of the journal policy strictness on the code re-execution rate. Finally, based on our results, we propose a set of recommendations for code dissemination aimed at researchers, journals, and repositories.
\end{abstract}


\section{Introduction}

Researchers increasingly publish their data and code to enable scientific transparency, reproducibility, reuse, or compliance with funding bodies, journals, and academic institutions~\cite{science_fane_ayris_hahnel_hrynaszkiewicz_baynes_farrell_2019}. Reusing data and code should propel new research and save researchers' time, but in practice, it is often easier to write new code than reuse old. Even attempting to reproduce previously published results using the same input data, computational steps, methods, and code has shown to be troublesome. Studies have reported a lack of research reproducibility~\cite{national2019reproducibility, baker_1500_2016} often caused by inadequate documentation, errors in the code, or missing files. 

Paradigms such as literate programming could help in making the shared research code more understandable, reusable, and reproducible. In literate programming, traditional source code is interspersed with explanations of its logic in a natural language~\cite{knuth1984literate}. The paradigm was encouraged for scientific computing and data science to facilitate reproducibility and transparency. However, in practice, researchers write code intending to obtain scientific insights, and there is often no incentive to structure and annotate it for reuse. As a result, the research code quickly becomes unusable or unintelligible after meeting its initial purpose~\cite{pasquier_sharing_2018}.

Though much of the code's intrinsic design will determine its longevity, its dissemination platform could also have a compelling influence~\cite{trisovic2021repository}. In particular, data and code repositories are some of the primary venues for sharing research materials. They aim to support researchers by creating general dissemination guidelines and descriptive metadata, but they cannot always prevent irreproducibility and code-rot due to the vast diversity of programming languages and complex computing processes. This is only aggravated as researchers generate and share new results and code at a rate higher than ever before. 

This paper presents a study that provides an insight into the programming literacy and reproducibility aspects of shared research code. The first premise of the study is to examine the properties of the shared datasets and research code. Information such as their size, content, presence of comments in the code, and documentation in the directory help us understand the current state of research code. By comparing the observed coding practices to the established best practices, we identify the existing weak points and areas of improvement for researchers writing code. Our content analysis gives us an insight into the storage needs and requirements for supporting files, such as documentation, images, or maps. The second premise of the study is to examine what happens when an external researcher retrieves and re-executes shared research code. In particular, we ask what the common errors are in executing this code and whether they can be solved with simple changes in the code. We explore if the code re-execution rates vary between different disciplines and other available features, and analyze the practices behind the best-performing ones. Finally,  based on the study's findings, we conclude with recommendations for disseminating research code for researchers, journals, and repositories.

\section{Background}\label{sec:bg}

Our study uses code deposited and shared at the Harvard Dataverse repository. The Dataverse project\footnote{http://dataverse.org} is an open-source data repository platform for  sharing, archiving, and citing research data. It is developed and maintained by the Harvard's Institute for Quantitative Social Sciences (IQSS) and a community of open source contributors. Currently, more than 60 institutions worldwide run Dataverse instances as their data repository, each hosting data generated by one or more institutions. 

Dataverse repositories allow researchers to deposit and share all research objects, including data, code, documentation, or any combination of these files. A bundle of these files associated with a published scientific result is called a replication package (or replication dataset).\footnote{It is often referred to as "replication data" in Dataverse repositories.} Researchers' code from replication packages usually operates on data to obtain the published result. For the Harvard Dataverse repository, replication packages are typically prepared and deposited by researchers themselves in an unmediated fashion (self-curated).

\begin{figure}
    \centering
    \includegraphics[width=\linewidth]{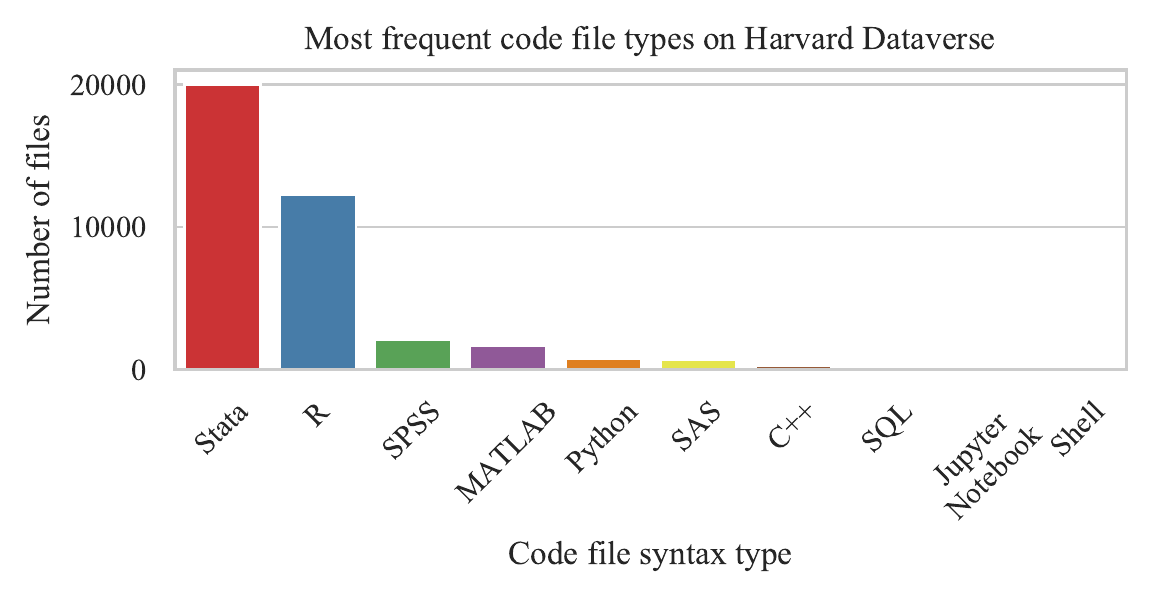}
    \caption{Most popular code file types on Harvard Dataverse (Oct, 2020). Of the top two, R is open source and free.}
    \label{fig:reprop}
\end{figure}

The most popular programming languages among the Harvard Dataverse repository users are Stata and R, as shown from the frequency of deposited code files in Figure~\ref{fig:reprop}. The two languages are often used in quantitative social science research. Their observed popularity can be attributed to Harvard Dataverse repository initially specializing in sharing social science research data. In the last five years, it has become a general-purpose, inter-disciplinary data repository. Stata is proprietary statistical software used in economics, sociology, political science, and health sciences. R is free and open-source software frequently used among statisticians and data analysts in the social sciences. Due to its popularity among academics and its open-source license, R is an ideal candidate for our study. It is currently ranked as the 13th most popular language in the TIOBE index\footnote{https://www.tiobe.com/tiobe-index/}. In the past, it was ranked as the most popular language~\cite{muenchen2012popularity} and has been rated among the top in the Kaggle Machine Learning \& Data Science Survey in the previous few years~\cite{kaggle}. R originated as an open-source and free version of S, a statistical command language that made programming accessible without the necessity of formal training. R is highly adaptable due to its extensible package system, which led to a surge of community-driven developments. Although the broad community development created potential for unsustainable code, methods for package standardization and quality control have been improving with the creation of RStudio, an integrated development environment (IDE) for R, and online communities like R-Hub and ROpenSci.

\section{Implementation and methods}

The R programming language is the main focus of our study due to its open-source license and popularity in scientific computing. We retrieve the content of 2109 publicly-available replication packages published from 2010 to July 2020 that contain 9078 R code files from the Harvard Dataverse repository. The Harvard Dataverse archives more than 40,000 datasets containing over 500,000 files at the time of writing. The rest of the datasets, over 65,000, are harvested from other federated repositories. For our analysis, we use only the deposited datasets (not harvested) due to the metadata differences across different repositories. Below, we elaborate on the study's implementation, workflow, and data collection.\footnote{Our study is loosely inspired by the effort undertaken by Chen~\cite{chen2018coding} in his undergraduate coursework, though our implementation, code-cleaning and analysis goals differ.}

\begin{figure}
    \centering
    \includegraphics[width=.9\linewidth]{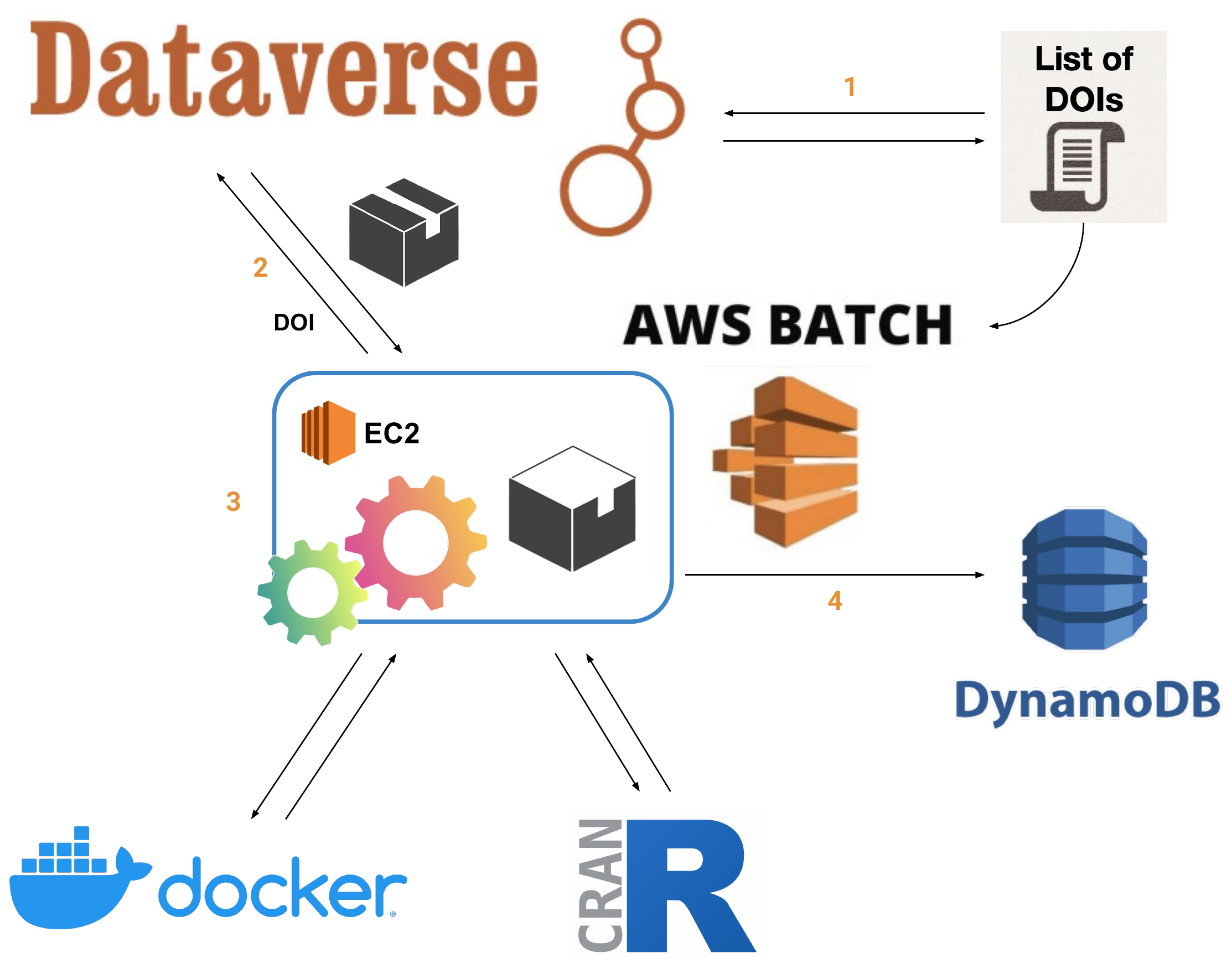}
    \caption{Implementation on the AWS Batch}
    \label{fig:batch}
\end{figure}

We use AWS Batch\footnote{https://aws.amazon.com/batch/} to parallelize the effort of retrieving and re-executing research code in each replication package. AWS Batch automatically provisions resources and optimizes the workload distribution while executing jobs without interactions with the end-user. All replication packages in the Harvard Dataverse repository are uniquely identified with a DOI (digital object identifier), and we start the analysis by retrieving the list of DOIs that contain R code (Figure~\ref{fig:batch}). 

\begin{enumerate}
    \item The DOI list is used to define the AWS jobs, which are then sent to the batch queue, waiting until resources become available for their execution.
    \item When a job leaves the queue, it instantiates a pre-installed Docker image that contains the necessary software pipeline to retrieve a replication package and execute its R code.
    \item Each job re-executes code from a single replication package using an EC2 instance with 16 vCPUs and 1024 GB of memory. 
    \item Finally, the results and information related to the re-execution are stored on DynamoDB.\footnote{https://aws.amazon.com/dynamodb/}
\end{enumerate}

The collected data, source code and complete instructions to reproduce our analysis are available online at Dataverse~\cite{study_data} and GitHub\footnote{https://github.com/atrisovic/dataverse-r-study} under MIT license.

\subsection{Data collection workflow}

For testing research code re-execution, we use a Docker image with pre-installed conda environment manager, R and Python software on Debian GNU/Linux 10. The image contains three independent R  environments, each with a different version of R interpreter and corresponding r-essentials, a bundle of approximately 200 most popular R packages for data science. In addition to the software, the image contains a custom-made workflow that conducts the study and collects data. The logic of the workflow is the following: 

\begin{enumerate}
    \item It downloads a replication package from the Harvard Dataverse repository. We verify and note if the file has correctly downloaded or if there was a checksum error. We collect data on the size and content of the replication package.
    \item We conduct an automatic code cleaning, scanning and correcting the code for some of the most common execution errors, such as hard-coded path variables (see the next section). Statistics on code files, such as the number of lines, libraries, and comments, are also collected.
    \item The workflow attempts to execute the researchers' code for an allocated period of one hour per file and five hours in total. The re-execution test is conducted with and without the code cleaning step, and the result (success, error, or time-limit exceeded) is recorded.
    \item The re-execution results and other collected data are passed to the backend database for analysis.
\end{enumerate}

Though a total of 2,170 replication packages contained R code and were visible through the Dataverse API, we successfully retrieved 2109 (97\%) of them. Some of these packages had restricted access and caused an 'authorization error' when we attempted to retrieve them. In other cases, files had obscure and erroneous encoding, which caused errors during the download.
Those were excluded from our study.

\subsection{Code cleaning}\label{sec:cleaning}

Our implementation of code cleaning aims to solve some of the most common re-execution errors. In particular, it removes absolute file paths, standardizes file encoding, and identifies and imports used libraries to set up a proper execution environment. The research code is modified to install the used library if it is not already present in the environment. The code cleaning approach is kept relatively simple to minimize the chance of 'breaking the code' or creating errors that were not previously there. Readers can learn more about the technical implementation of code cleaning in Appendix~\ref{cleaning}.

\section{Results and discussion}

We define ten research questions to provide a framework for the study. The first group of questions revolves around coding practices (RQ 1-3), while the other around the automated code re-execution (RQ 4-10).

\rquest{RQ 1. What are the basic properties of a replication package in terms of its size and content?}

\begin{figure}[h]
     \centering
     \begin{subfigure}[]{0.48\textwidth}
         \centering
         \includegraphics[width=.95\textwidth]{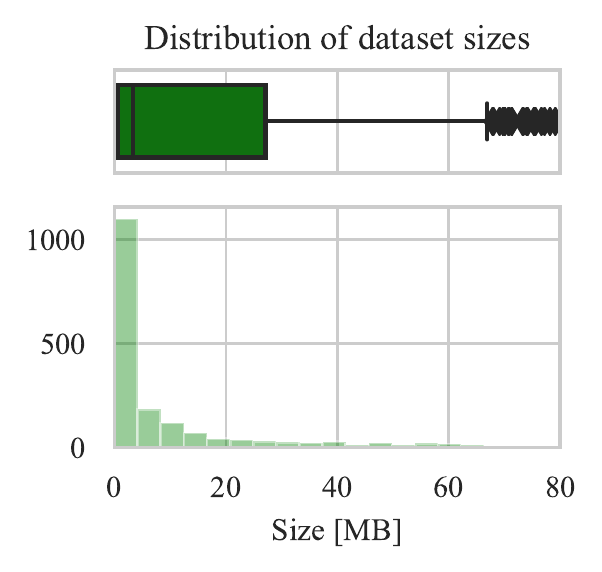}
     \end{subfigure}
     \begin{subfigure}[]{0.48\textwidth}
         \centering
         \includegraphics[width=.95\textwidth]{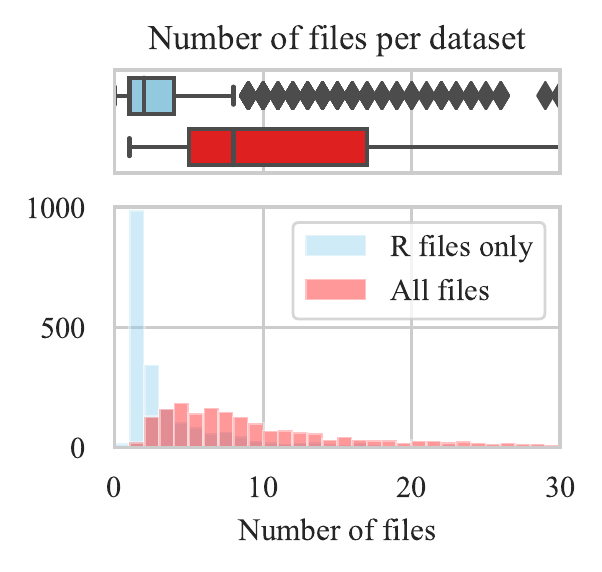}
     \end{subfigure}
        \caption{Dataset sizes and file counts}
        \label{fig:size}
\end{figure}

Our first research question focuses on the basic dataset properties, such as its size and content. The average size of a dataset is 92 MB (with a median of 3.2 MB), while the average number of files in a dataset is 17 files (the median is 8). Even though it may seem that there is a large variety between datasets, by looking at the distributions (Figure~\ref{fig:size}), we observe that most of the datasets amount to less than 10 MB and contain less than 15 files.

Analyzing the content of replication packages, we find that about 40\% of them (669 out of 2,091) contain code in other programming languages (i.e., not R). Out of 2091 datasets, 620 contained Stata code (.do files), 46 had Python code (.py files), and 9, 7, and 6 had SAS, C++, and MATLAB code files, respectively. The presence of different programming languages can be interpreted in multiple ways. It might be that a dataset resulted in a collaboration of members who preferred different languages, i.e., one used R and another Stata. Alternatively, it may be that different analysis steps are seamlessly done in different languages, for example, data wrangling in R and visualization in Python. However, using multiple programming languages may hinder reproducibility, as an external user would need to obtain all necessary software to re-execute the analysis. Therefore, in the re-execution stage of our study, it is reasonable to expect that replication packages with only R code would perform better than those with multiple programming languages (where R code might depend on the successful execution of the code in other languages).

The use of R markdown and Rnw have been encouraged to facilitate result communication and transparency~\cite{baumer2014r}. R markdown (RMD) files combine formatted plain text and R code that provide a narration of research results and facilitate their reproducibility. Ideally, a single command can execute the code in an R markdown file to reproduce reported results. Similarly, Rnw (or Sweave) files combine content such as R code, outputs, and graphics within a document. We observe that only a small fraction of datasets contain R markdown (3.11\%) and Rnw files (0.24\%), meaning that to date few researchers have employed these methods.

Last, we observe that 91\% of the files are encoded in ASCII and about 5\% in UTF-8. The rest use other encodings, with ISO-885901 and Windows-1252 being the most popular (about 3.5\% together). In the code cleaning step, all non-ASCII code files (692 out of 8173) were converted into ASCII to reduce the chance of encoding error. In principle, less popular encoding formats are known to sometimes cause problems, so using ASCII and UTF-8 encoding is often advised~\cite{world2014best}.

\rquest{RQ 2. Does the research code comply with literate programming and software best practices?}

There is a surge of literature on best coding practices and literate programming~\cite{martin_clean_2009, sandve_ten_2013, mcconnell2004code, thomas2019pragmatic, hyde2020write} meant to help developers create quality code. One can achieve higher productivity, easier code reuse, and extensibility by following the guidelines, which are typically general and language agnostic. In this research question, we aim to assess the use of best practices and programming literacy in the following three aspects: meaningful file and variable naming, presence of comments and documentation, and code modularity through the use of functions and classes.

\begin{figure}[h]
     \centering
     \begin{subfigure}[]{0.48\textwidth}
         \centering
         \includegraphics[width=.95\textwidth]{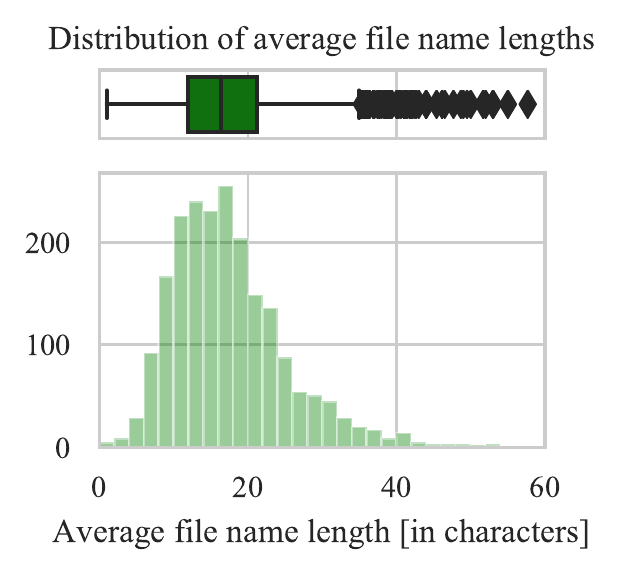}
         \caption{}
         \label{fig:file_name}
     \end{subfigure}
     \begin{subfigure}[]{0.48\textwidth}
         \centering
         \includegraphics[width=.9\textwidth]{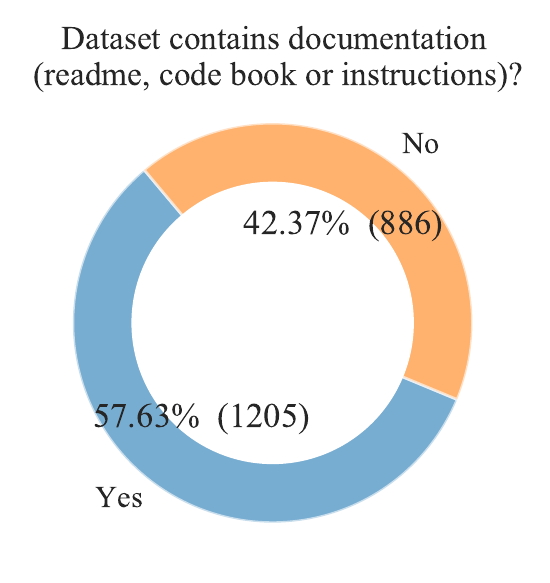}
         \caption{}
         \label{fig:doc}
     \end{subfigure}
        \caption{File name lengths and presence of documentation}
\end{figure}

Best practices include creating descriptive file names and documentation. Indeed, if file names are long and descriptive, it is more likely that they will be understandable to an external researcher. The same goes for additional documentation within the dataset. We observe that the average filename length is 17 characters without the file extension. The filename length distribution (Figure~\ref{fig:file_name}) shows that most file lengths are between 10 and 20 characters. However, we note that about a third or 32\% (669) of file names contain a ‘space’ character, which is discouraged as it may hinder its manipulation when working from the command line. We also searched for a documentation file, or a file that contains “readme,” “codebook,” “documentation,” “guide,” and “instruction” in its name, and found it in 57\% of the datasets (Figure~\ref{fig:doc}). The authors may have also adopted a different convention to name their documentation material. Therefore, we can conclude that the majority of authors upload some form of documentation alongside their code.

\begin{figure}[h]
     \centering
     \begin{subfigure}[]{0.48\textwidth}
         \centering
         \includegraphics[width=.95\textwidth]{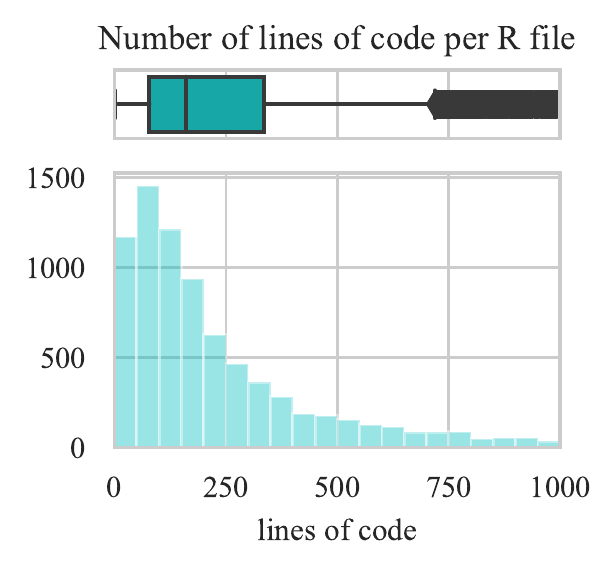}
         \caption{}
         \label{fig:lines}
     \end{subfigure}
     \begin{subfigure}[]{0.48\textwidth}
         \centering
         \includegraphics[width=.95\textwidth]{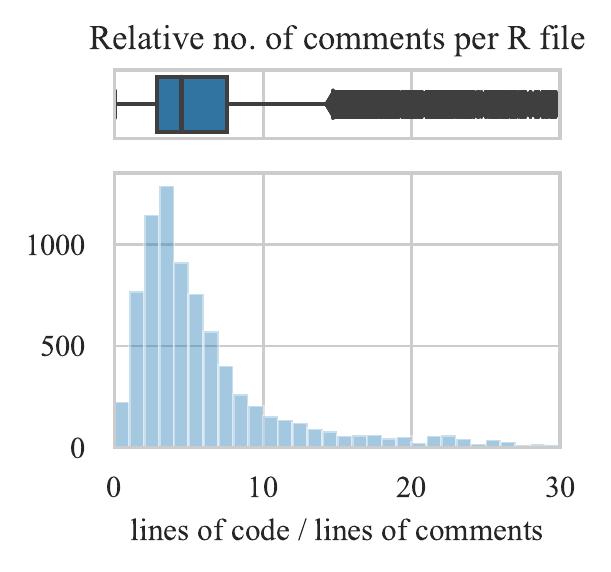}
         \caption{}
         \label{fig:comm}
     \end{subfigure}
        \caption{Number of code line and relative number of comments.}
\end{figure}

Further, we examine the code of the 8875 R files included in this study. The average number of code lines per file is 312 (the median is 160) (Figure~\ref{fig:lines}). Considering that there are typically 2 R files per dataset (median, mean is 4), we can approximate that behind each published dataset lies about 320 R code lines. 

Comments are a frequent part of code that can document its processes or provide other useful information. However, sometimes they can be redundant or even misleading to a reuser. It is good practice to minimize the number of comments not to clutter the code and replace them with intuitive names for functions and variables~\cite{martin_clean_2009}. To learn about commenting practices in the research code, we measure the ratio of code lines and comment lines for each R file. The median value of 4.5 (average is 7) can be interpreted as one line of comments documenting 4 or 5 lines of code (Figure~\ref{fig:comm}). In other words, we observe that comments comprise about 20\% of the shared code. Though the optimal amount of comments depends on the use case, a reasonable amount is about 10\%, meaning that the code from Harvard Dataverse is on average commented twice as much.

According to IBM studies, intuitive variable naming contributes more to code readability than comments, or for that matter, any other factor~\cite{mcconnell2004code}. The primary purpose of variable naming is to describe its use and content, therefore, they should not be single characters or acronyms but words or phrases. For this study, we extracted variable names from the code using the built-in R function \texttt{ls()}. Out of 3070 R files, we find that 621 use variables that are one or two characters long. However, the average length of variable names is 10, which is a positive finding as such a name could contain one, two, or more English words and be sufficiently descriptive to a reuser.

Modular programming is an approach where the code is divided into sections or modules that execute one aspect of its functionality. Each module can then be debugged, understood, and reused independently. In R programming, these modules can be implemented as functions or classes. We count the occurrences of user-defined functions and classes in R files to learn how researchers structure their code. 

Out of 8875 R files, 2934 files have either functions or classes. Applying a relative number of modules per lines of code, we can estimate that one function on average contains 82 lines of code (mode is 55). According to a synthesis of interviews with top software engineers, a function should include about 10 lines of code~\cite{martin_clean_2009}. However, as noted in Section~\ref{sec:bg}, R behaves like a command language in many ways, which does not inherently require users to create modules, such as functions and classes. Along these same lines, R programmers do not usually refer to their code collectively as a program but rather as a script, and it is often not written with reuse in mind. 

\rquest{RQ 3. What are the most used R libraries?}

\begin{figure}
    \centering
    \includegraphics{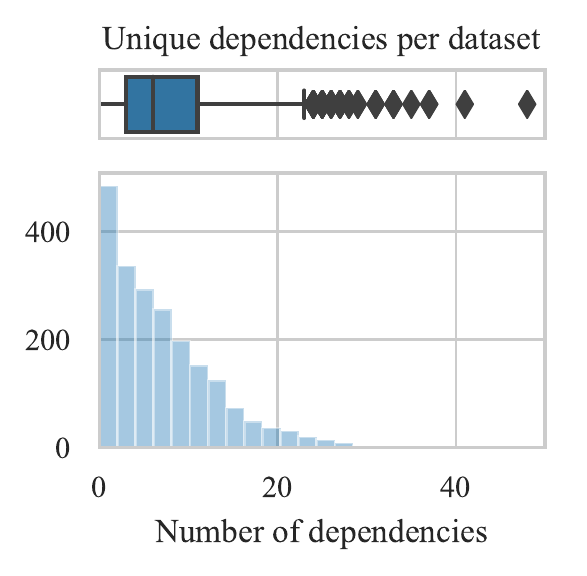}
    \caption{Distribution of unique dependencies per dataset.}
    \label{fig:libs}
\end{figure}

The number of code dependencies affects the chances of reusing the code, as all dependencies (of adequate versions) need to be present for its successful execution. Therefore, a higher number of dependencies can lower the chances of their successful installation and ultimately code re-execution and reuse. We find that most datasets explicitly depend on up to 10 external libraries  (Figure~\ref{fig:libs}) with individual R files requiring an average of 4.3 external dependencies (i.e., R libraries).\footnote{The dependencies in code are detected with “library”, “install.packages” and “require”.}

\begin{figure}[h]
    \centering
    \includegraphics[width=\linewidth]{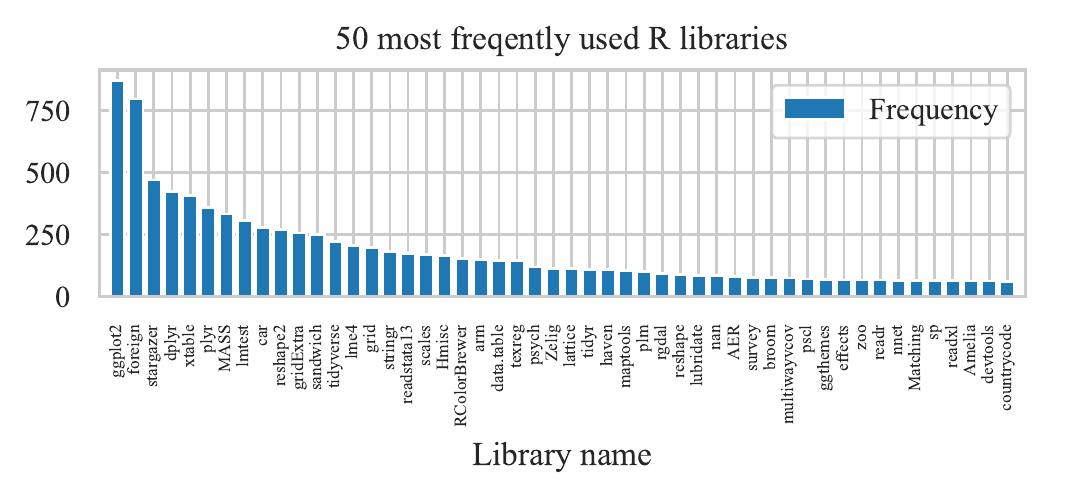}
    \caption{Most frequently used R libraries.}
    \label{fig:pop}
\end{figure}

The list of used libraries provides insight into the goals of research code (Figure~\ref{fig:pop}). Across all of the datasets, the most frequently used library is \texttt{ggplot2} for plotting, indicating that the most common task is data visualization. Another notable library is \texttt{xtable}, which offers functions for managing and displaying data in a tabular form, and similarly provides data visualization. Many libraries among the top ten are used to import and manage data, such as \texttt{foreign}, \texttt{dplyr}, \texttt{plyr} and \texttt{reshape2}. Finally, some of them are used for statistical analysis, like \texttt{stargazer}, \texttt{MASS}, \texttt{lmetest} and \texttt{car}. These libraries represent the core activities in R: managing and formatting data, analyzing it, and producing visualizations and tables to communicate results. The preservation of these packages is, therefore, crucial for reproducibility efforts.

The infrequent usage and absence of libraries also tell us what researchers are not doing in their projects. In particular, libraries that are used for code testing, such as \texttt{runit}, \texttt{testthat}, \texttt{tinytest} and \texttt{unitizer}, were not present. Although these libraries are primarily used to test other libraries, they could also confirm that data analysis code works as expected. For example, tests of user-defined functions, such as data import or figure rendering, could be implemented using these libraries. Another approach that can aid in result validation and facilitate reproducibility is computational provenance~\cite{davidson2008provenance, pasquier2017if}. It refers to tracking data transformations with specialized R libraries, such as \texttt{provR}, \texttt{provenance}, \texttt{RDTlite}, \texttt{provTraceR}. However, in our study, we have not found a single use of these provenance libraries. In addition, libraries for runtime environment and workflow management are also notably absent. Libraries such as \texttt{packrat} and \texttt{pacman} aid in the runtime environment management, and workflow libraries, such as \texttt{workflowR}, \texttt{workflows}, and \texttt{drake} provide explicit methods for reproducibility and workflow optimization (e.g., via caching and resource scaling). None of these were detected in our dataset. Though we can conclude that these approaches are not currently intuitive for the researchers, encouraging their use could significantly improve research reproducibility and reuse.

Finally, we look for configuration files, which are used to build a runtime environment and install code dependencies. One of the common examples of a Python configuration file is \texttt{requirements.txt}, though several other options exist. In some conventions, a configuration file for R should be named \texttt{install.R}. We have not found a single \texttt{install.R} among the analyzed datasets, but we found similar files such as: \texttt{installrequirements.R}, \texttt{000install.R}, \texttt{packageinstallation.R} or \texttt{postinstall}. We can see that there is currently no ubiquitously-used convention, though the R community would benefit from establishing it.

\rquest{RQ 4. What is the code re-execution rate?}

We re-executed R code from each of the replication packages using three R software versions, R 3.2, R 3.6, and R 4.0, in a clean environment. The possible re-execution outcomes for each file can be a "success", an error, and a time limit exceeded (TLE). TLE occurs when the time allocated for file re-execution is exceeded. We allocated up to 5 hours of execution time to each replication package, and within that time, we allocated up to 1 hour to each R file. The execution time may include installing libraries or external data download if those are specified in the code. The replication packages that have resulted in TLE were excluded from the study, as they may have eventually executed properly with more time (some would take days or weeks). To analyze the success rate of the analysis, we interpret and combine the results from the three R versions in the following manner (illustrated in Table~\ref{tab:comb}):

\begin{enumerate}
    \item If there is a "success" one or more times, we consider the re-execution to be successful. In practice, this means that we have identified a version of R able to re-execute a given R file.
    \item If we have one or more "TLE" and no "success", the combined result is a TLE. The file is then excluded to avoid misclassifying a script that may have executed if given more time. 
    \item Finally, if we have an "error" 3 times, we consider the combined results to be an error.
\end{enumerate}

\begin{table}[h]
\centering
\begin{tabular}{llll}
\toprule
\multicolumn{3}{l}{\textbf{Outcomes}}        & \textbf{Combined result} \\
\midrule
Success & Anything    & Anything    & \textbf{Success}         \\
TLE     & Not Success & Not Success & \textbf{TLE}             \\
Error   & Error       & Error       & \textbf{Error}          \\
\bottomrule
\end{tabular}
\caption{Obtaining combined re-execution result per R file from results using three versions of R software. TLE means "time limit exceeded".}\label{tab:comb}
\end{table}

We re-execute R code in two different runs, with and without code cleaning, using in each three different R software versions. We note that while 9078 R files were detected in 2109 datasets, not all of them got assigned a result. Sometimes R files exceeded the allocated time, leaving no time for the rest of the files to execute. When we combine the results based on the Table~\ref{tab:comb}, we get the following results in each of the runs:

\begin{table}[h]
\centering
\begin{tabular}{lccc}
\toprule
                        & \textbf{\begin{tabular}[c]{@{}c@{}}Without \\ code cleaning\end{tabular}} & \textbf{\begin{tabular}[c]{@{}c@{}}With \\ code cleaning\end{tabular}} & \textbf{Best of both} \\
                        \midrule
\textbf{Success rate}   & 25\%                                                                      & 40\%                                                                   & 56\%                  \\
\textbf{Success}        & 952                                                                       & 1472                                                                   & 1581                  \\
\textbf{Error}          & 2878                                                                      & 2223                                                                   & 1238                  \\
\textbf{TLE}            & 3829                                                                      & 3719                                                                   & 5790                  \\
\textbf{Total files}    & 7659                                                                      & 7414                                                                   & 8609                  \\
\textbf{Total datasets} & 2071                                                                      & 2085                                                                   & 2109                 \\
\bottomrule
\end{tabular}
\end{table}

Going forward, we consider the results with code cleaning as primary and further analyze them unless different is stated.

\rquest{RQ 5. Can automatic code cleaning with small changes to the code aid in its re-execution?} 

To determine the effects of the code cleaning algorithm (described in Section~\ref{sec:cleaning}), we first re-execute original researchers' code in a clean environment. Second, we re-execute the code after it was modified in the code-cleaning step. We find an increase in the success rate for all R versions, with a total increase of about 10\% in the combined results.\footnote{The difference of 10\% was obtained by comparing the files that have explicit errors and successes and excluding the ones with TLE values.}

\begin{figure}[h]
    \centering
    \includegraphics[width=\linewidth]{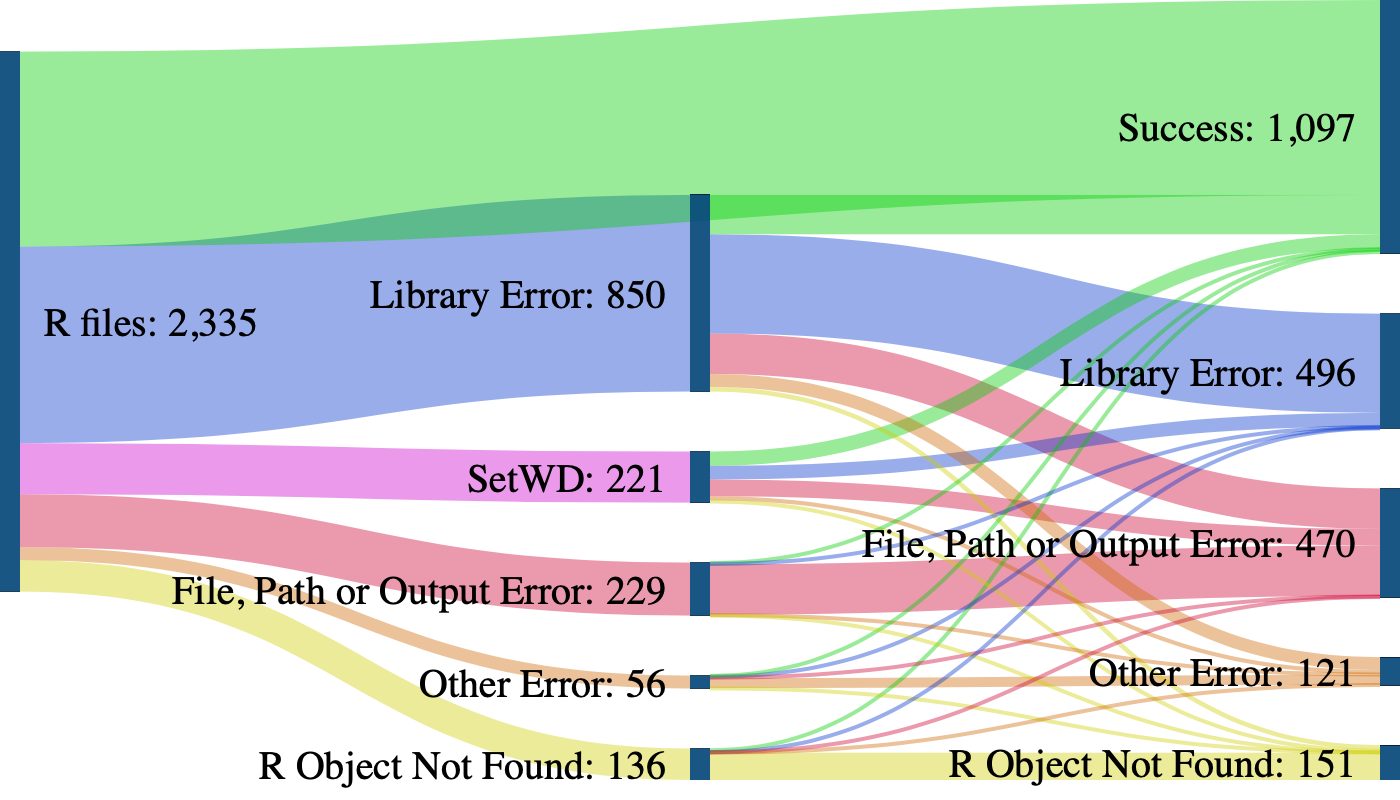}
    \caption{Success rate and errors before and after code cleaning. To objectively determine the effects of code cleaning, we subset the results that have explicit "successes" and errors while excluding the ones with TLE values as the outcome. As a result, the count of files in this figure is lower than the total count.}
    \label{fig:sankey}
\end{figure}

Looking at the breakdown of coding errors in Figure~\ref{fig:sankey}, we see that code cleaning can successfully address some errors. In particular, it fixed all errors related to the command \texttt{setwd} that sets a working directory and is commonly used in R. Another significant jump in the re-execution rate results from resolving the errors that relate to the used libraries. Our code cleaning algorithm does not pre-install the detected libraries but instead modifies the code to check if a required library is present and installs it if it is not (see Appendix~\ref{cleaning}). 

Most code files had other, more complex errors after code cleaning resolved the initial ones. For example, other library errors appeared if a library was not installed or was incompatible due to its version. Such an outcome demonstrates the need to capture the R software and dependency versions required for reuse. File, path, and output errors often appeared if the directory structure was inaccurate or if the output file was not saved. "R object not found" error occurs when using a variable that does not exist. While it is hard to pin the cause of this error precisely, it is often related to missing files or incomplete code. 
Due to the increased success rate with code cleaning, we note that many common errors could be avoided. There were no cases of code cleaning "breaking" the previously successful code, meaning that a simple code cleaning algorithm, such as this one, can improve code re-execution. Based on our results, we give recommendations in Section~\ref{sec:rec}. 

\rquest{RQ 6. Are code files designed to be independent of each other or part of a workflow? }

\begin{figure}[h]
    \centering
    \includegraphics[width=\linewidth]{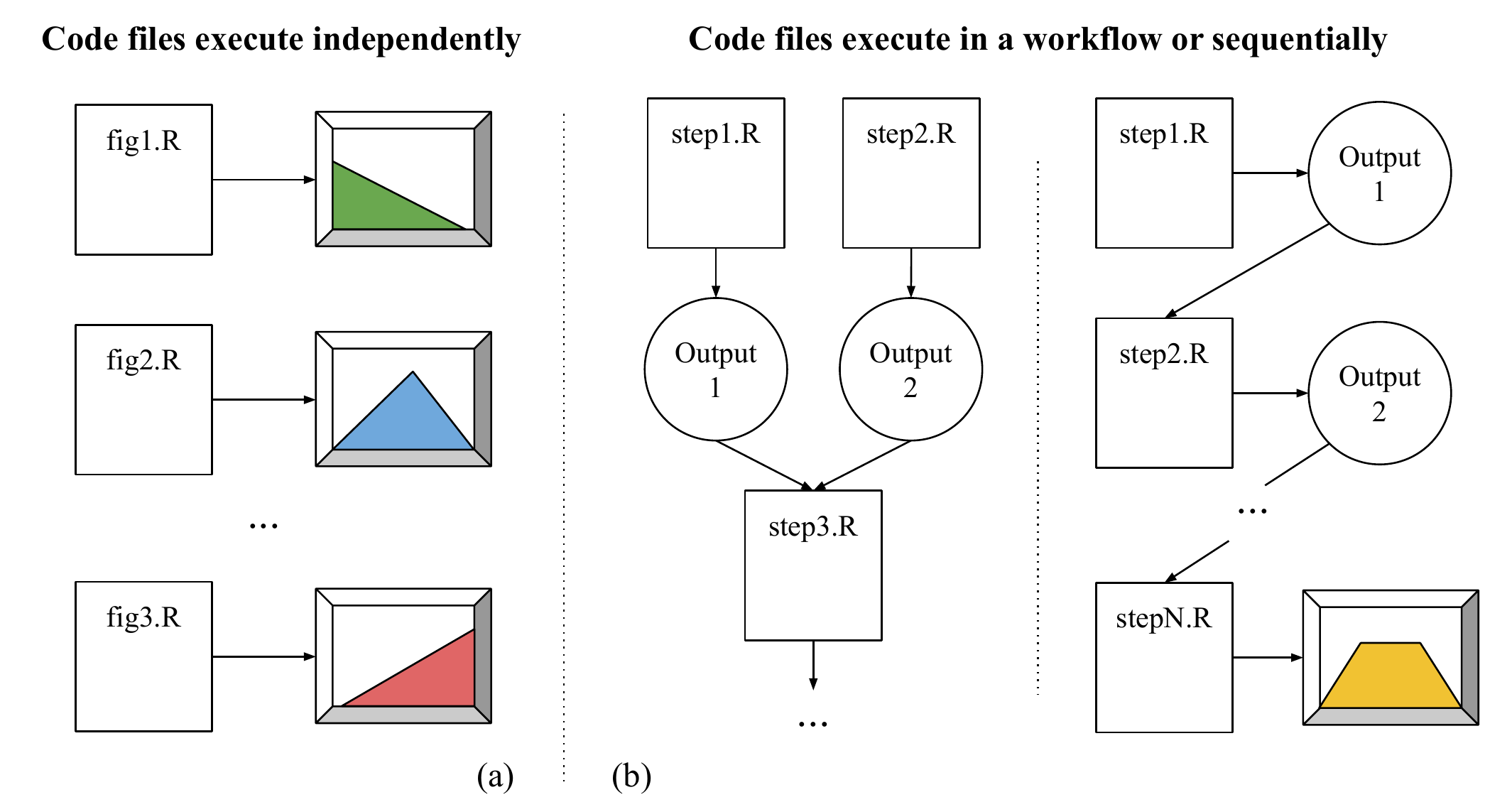}
    \caption{Types of workflows in research analyses.}
    \label{fig:wflowst}
\end{figure}

R files in many datasets are designed to produce output independently of each other  (Figure~\ref{fig:wflowst}a), while some are structured in a workflow  (Figure~\ref{fig:wflowst}b), meaning that the files need to be executed in a specific order to produce a result.  Due to the wide variety of file naming conventions, we are unable to detect the order in which the files should be executed. As a result, we may run the first step of the workflow last in the worst case, meaning that only one file (the first step) will run successfully in our re-execution study. 

To examine the nature of R analysis, we aggregate the collected re-execution results in the following fashion. If there are one or more files that successfully re-executed in a dataset, we mark that dataset as 'success'. A dataset that only contains errors is marked as 'error', and datasets with TLE values are removed. In these aggregated results (dataset-level), 45\% of the datasets (648 out of 1447) have at least one automatically re-executable R file. There is no drastic difference between the file-level success rate (40\%) and the dataset-level success rate (45\%), suggesting that the majority of files in a dataset are meant to run independently. However, the success rate would likely be better had we known the execution order. 

If we exclude all datasets that contain code in other programming languages, the file-level success rate is 38\% (out of 2483 files), and the dataset-level success rate is 45\% (out of 928 datasets). These ratios are comparable to the ones in the whole dataset (40\% on file-level and 45\% on dataset-level), meaning that "other code" does not significantly change the re-execution success rate. In other words, we would expect to see a lower success rate if an R file depends on the execution of the code in other languages. Such a result corroborates the assumption that R files were likely designed to be re-executed independently in most cases.

\rquest{RQ 7. What is the success rate in the datasets belonging to journal Dataverse collections?}

\begin{figure}
    \centering
    \includegraphics[width=\linewidth]{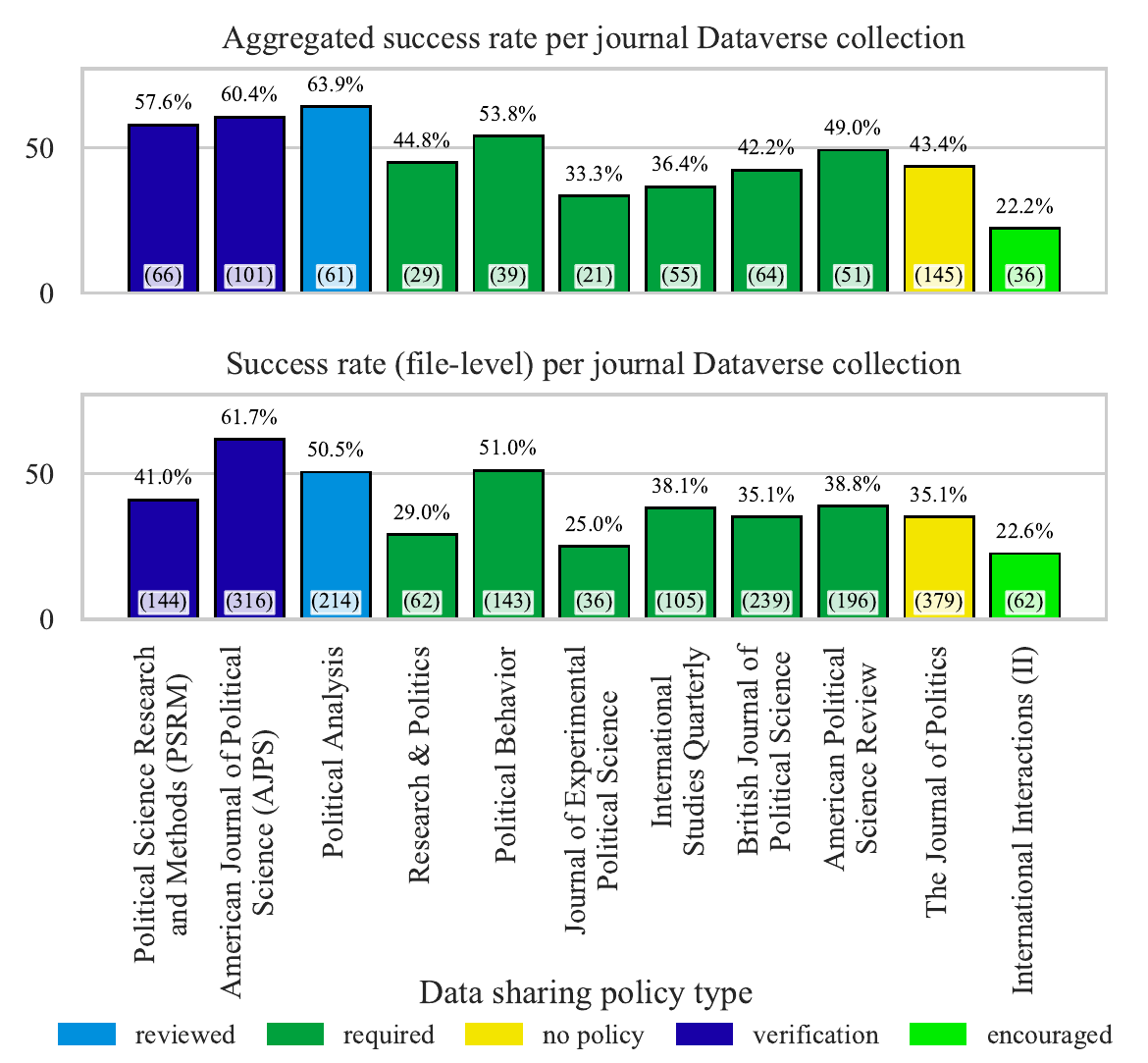}
    \caption{Re-execution success rate per journal Dataverse collection. In the brackets are the number of datasets and the number of R files, respectively.}
    \label{fig:journal}
\end{figure}

More than 80 academic journals have their dedicated data collections within the Harvard Dataverse repository to support data and code sharing as supplementary material to a publication. Most of these journals require or encourage researchers to release their data, code, and other material upon publication to enable research verification and reproducibility. By selecting the datasets linked to a journal, we find a slightly higher than average re-execution rate (42\% and aggregated 47\% instead of 40\% and 45\%). We examine the data further to see if a journal data sharing policy influences its re-execution rate. 

We survey data sharing policies for a selection of journals and classify them into five categories according to whether data sharing is: encouraged, required, reviewed, verified, or there is no policy. We analyze only the journals with more than 30 datasets in their Dataverse collections. Figure~\ref{fig:journal} incorporates the survey and the re-execution results. "No-policy" means that journals do not mandate the release of datasets. "Encouraged" means that journals suggest to authors to make their datasets available. "Required" journals mandate that authors make their dataset available. "Reviewed" journals make datasets part of their review process and ensure that it plays a role in the acceptance decision. For example, the journal Political Analysis (PA) provides detailed instruction on what should be made available in a dataset and conducts "completeness reviews" to ensure published datasets meet those requirements. Finally, "verified" means the journals ensure that the datasets enable reproducing the results presented in a paper. For example, the American Journal of Political Science (AJPS) requires authors to provide all the research material necessary to support the paper claims. Upon acceptance, the research material submitted by authors is verified to ensure that they produce the reported results~\cite{vines_mandated_2013, crosas_data_2018}. From Figure~\ref{fig:journal} we see that the journals with the strictest policies (Political Science Research and Methods, AJPS, and PA) have the highest re-execution rates. Therefore, our results suggest that the strictness of the data sharing policy is positively correlated to the re-execution rate of code files.

\begin{table}[h]
\centering
\begin{tabularx}{\textwidth}{lcccccccc}
\toprule
\textbf{Publication year} & 2010 & 2014 & 2015 & 2016 & 2017 & 2018 & 2019 & 2020 \\
\midrule
\textbf{Dataset count}    & 1    & 2    & 90   & 284  & 335  & 493  & 555  & 241 \\
\bottomrule
\end{tabularx}
\caption{Dataset publication date.}\label{tab:dset}
\end{table}

\rquest{RQ 8. Is the re-execution rate correlated with the R version considering that the datasets were released in a span of several years?}

R libraries are not often developed with backward-compatibility, meaning that using a different version from the one used originally, might cause errors when re-executing the code. Furthermore, different versions of R software might not be compatible with different versions of the libraries. The datasets in our study were published from 2010 to July 2020 (Table~\ref{tab:dset}), which gives us a unique perspective in exploring the potential of backward-compatibility in R.

\begin{table}[h]
\centering
\begin{tabular}{ll}
\toprule
\textbf{R version} & \textbf{Release date} \\
\midrule
R 3.2.1   & June, 2015   \\
R 3.6.0   & April, 2019  \\
R 4.0.1   & June, 2020  \\
\bottomrule
\end{tabular}
\caption{Release date of used R versions.}\label{tab:year}
\end{table}

\begin{figure}[h]
    \centering
    \includegraphics[width=\linewidth]{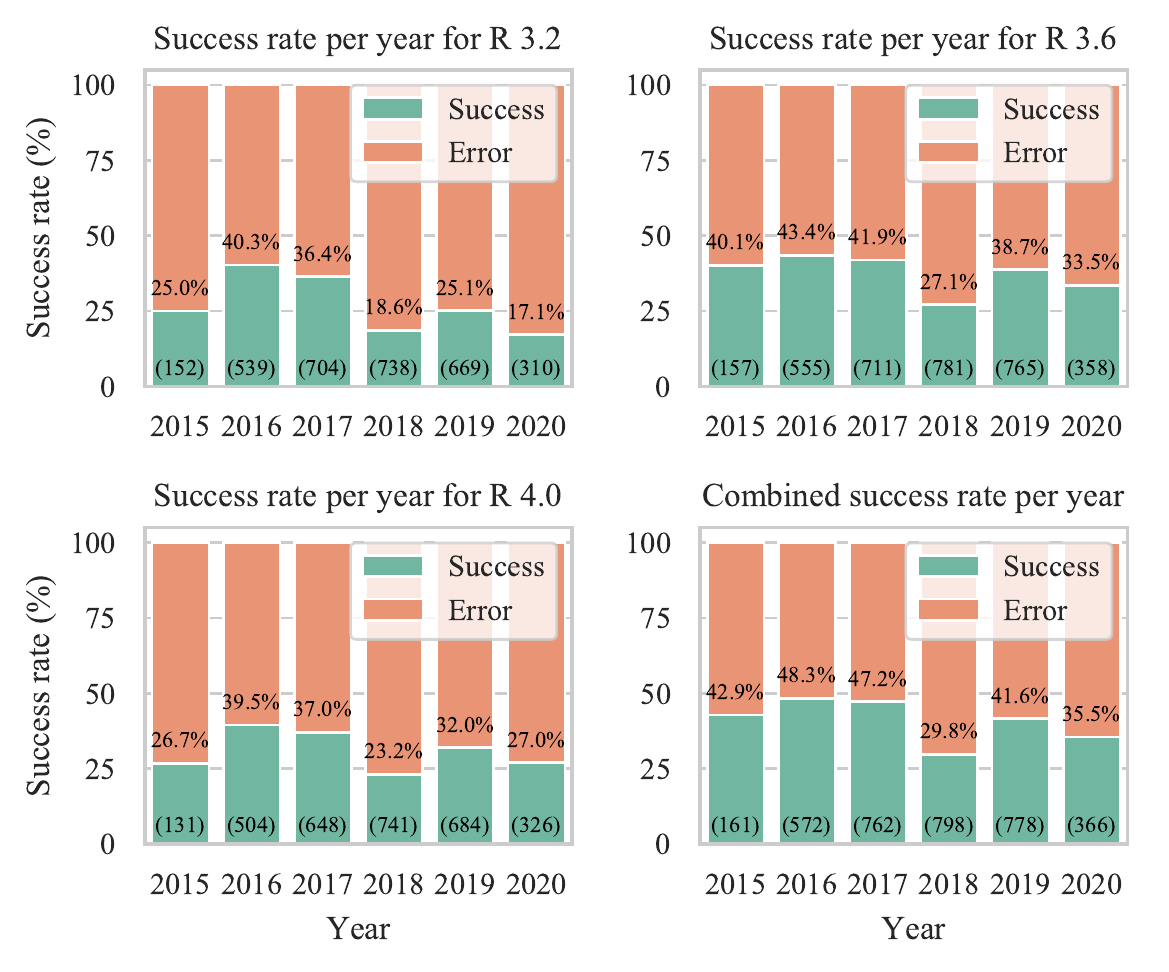}
    \caption{Re-execution success rates per year per R software version.}
    \label{fig:year}
\end{figure}

Considering the release years of used R versions (Table~\ref{tab:year}), we examine the correlation in the success rates between them and the release year of the replication package (Figure~\ref{fig:year}). We find that R 3.2, released in 2015, performed best with the replication packages released in 2016 and 2017. Such a result is expected because these replication packages were likely developed in 2015, 2016, and 2017 when R 3.2 was frequently used. We also see that it has a lower success rate in recent years. We observe that R 3.6 has the highest success rate per year. This version of R likely had some backward compatibility with older R subversions, which explains its high success rate in 2016 and 2017. Lastly, R 4.0 is a recent version representing a significant change in the software, which explains its generally low success rate. Because R 4.0 was released in summer 2020, likely none of the examined replication packages originally used that version of the software. We observe some evidence of backward compatibility. However, we do not find a significant correlation between the R version and the release year of a replication package.
A potential cause may be the use of incompatible library versions in our re-execution step as the R software automatically installs the latest version of a library. In any case, our result highlights that the execution environment evolves over time and that additional effort is needed to ensure that one could successfully recreate it for reuse.

\rquest{RQ 9. What is the success rate per research field? }

\begin{figure}[h]
    \centering
    \includegraphics[width=\linewidth]{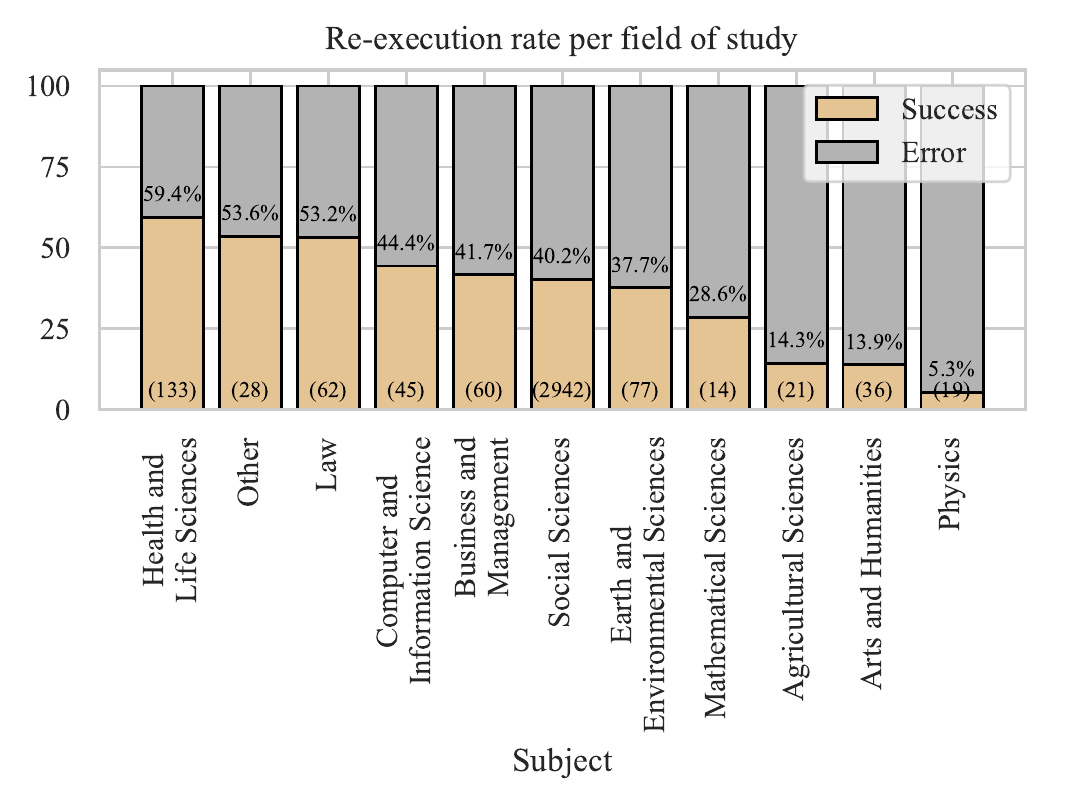}
    \caption{Success rate per research field}
    \label{fig:subjects}
\end{figure}

The Harvard Dataverse repository was initially geared toward social science, but it has since become a multi-disciplinary research data repository. Still, most of the datasets are labeled 'social science,' though some have multiple subject labels. To avoid sorting the same dataset into multiple fields, if a dataset was labeled both "social science" and "law," we would keep only the latter. In other words, we favored a more specific field (such as "law") and chose it over a general one (like "social science") when that was possible.

The re-execution rates per field of study are shown in Figure~\ref{fig:subjects}. The highest re-execution rates were observed for the health and life sciences. It may be that medical-related fields have a stronger level of proof embedded in their research culture. Physics had the lowest re-execution rate, though due to the low sample size and the fact that Dataverse is probably not the repository of choice in physics, we cannot draw conclusions about the field. Similarly, we cannot draw conclusions for many of the other fields due to the low sample size (see the number of R files in the brackets), and therefore, the results should not be generalized.

\rquest{RQ 10. How does the re-execution rate relate to research reproducibility?}

Research is reproducible if the shared data and code produce the same output as the reported one. Code re-execution is one of the essential aspects of its quality and a prerequisite for computational reproducibility. However, even when the code re-executes successfully in our study, it does not mean that it produced the reported results. To access how the code re-execution relates to research reproducibility, we select a random sample of three datasets where all files were executed successfully and attempt to compare its outputs to the reported ones. There are 127 datasets where all R files re-executed with success (Table~\ref{tab:outcomes}).

\begin{table}[h]
\centering
\begin{tabular}{lc}
\toprule
\textbf{Outcomes}         & \textbf{Dataset count} \\
\midrule
Only success          & 127            \\
Only error            & 654            \\
Only TLE              & 684            \\
Success \& error      & 167            \\
Success \& TLE        & 215            \\
Error \& TLE          & 143            \\
Success, error \& TLE & 114            \\
\midrule
\textbf{Total}        & 2085          \\
\bottomrule
\end{tabular}
\caption{Re-execution results combination per dataset. For example, there are 215 datasets that contain only 'success' and 'TLE' as outputs.}\label{tab:outcomes}
\end{table}

The first dataset from the random sample is a replication package linked to a published paper at the Journal of Experimental Political Science~\cite{DVN/SWV9GJ_2017}. It contains three R files and a Readme, among other files. The Readme explains that each of the R files represents a separate study and that the code logs are available within the dataset. Comparing the logs before and after code re-execution would be a good indication of its success. After re-executing two of the R files, we find that the log files are almost identical and contain identical tables. The recreated third log file nearly matched the original, but there were occasional discrepancies in some of the decimal digits (though the outputs were in the same order of magnitude). Re-executing the third R file produced a warning that the library \texttt{SDMTools} is not available for R 3.6, which may have caused the discrepancy.

The second dataset from the random sample is a replication package linked to a paper published at Research \& Politics~\cite{DVN/Z02C8Y_2019}. It has a single R file and a Readme explaining that the script produces a correlation plot. According to our results, it should be re-executed with R 3.2. The R file prints two correlation coefficients, but it is unable to save the plot in the Docker container. It prints out a warning pointing to a broken Linux dependency.\footnote{https://askubuntu.com/questions/1094742/libjpeg-turbo8-dev-broken-package-dependencies-16-04} 

The final dataset from our sample follows a paper published at the Review of Economics and Statistics~\cite{DVN/YT45AO_2016}. It contains two R files and a Readme. One of the R files contains only functions, while the other calls the functions file. Running the main R file prints a series of numbers, which are estimates and probabilities specified in the document. However, the pop-up plotting functions are suppressed due to the re-execution in the Docker container. The code does not give errors or warnings.

While successful code re-execution is not a sufficient measure for reproducibility, our sample suggests that it might be a good indicator that computational reproducibility will be successful.

\section{Limitations of the study}

Though this study is framed around pre-defined research questions, it has certain limitations in answering them due to its automatized and large-scale nature, which cannot detect the nuances in research code. Though code re-execution is a  critical aspect of reproducibility, it is not an equivalent indicator, as code might produce results that differ from the reported ones. Therefore, we would get a "false positive" as a result in our study. However, it is likely that small changes in the re-execution would also result in a higher success rate. In some cases, the code would not crash if it was re-executed in RStudio or if it had an available directory for saving outputs. All things considered, we note that our automated study has a comparable success rate to the reported manual reproducibility studies~\cite{stodden_empirical_2018, chang2015economics}, which gives strength to the overall significance of our results. Though conducting a reproducibility study with human intervention would result in more sophisticated findings, it is labor-intensive on a large scale. We should strive toward enabling reproducibility studies with automation while using existing standards like machine-readable FAIR data~\cite{wilkinson_fair_2016} and code~\cite{lamprecht2020towards, jimenez2017four}, and this paper provides recommendations toward that goal.

Even though it may appear that we should have tested the code with a higher number of R software versions, we believe that our results would not have been drastically different. Indeed this is a limitation of all R software versions, as each would by default try to install the latest version of R libraries. As a result, the research code would fail if the R version and the library version were not compatible. Even if the library was successfully installed, it might not work as expected if the author used an earlier version.

Some of the limitations were imposed by the number of resources we had on the AWS cloud. For instance, there is a large count of TLE values in our results even though we set the time limitation to be 1 hour per R file and 5 hours per dataset. The main cause of depleting the AWS resources was a choice of a large EC2 instance for the re-execution. In hindsight, these instances were excessive for this type of study, and we should have used small to medium-sized instances. 

Finally, one might consider a limitation that we use datasets from a single source. While code repositories such as GitHub are used for software development, temporary projects, research, education, and more, the datasets published on Harvard Dataverse are primarily created for research purposes. Furthermore, the data curation team at Harvard vets all published datasets to maintain its data and research focus. Therefore, this study provides an insight into researchers’ coding practices in R, and for that reason, it required a research-focused repository such as Dataverse.

\section{Best practices and recommendations}\label{sec:rec}

In this study, we saw that many errors in research code could be prevented with good practices. Extensive guides on code practices have been published~\cite{sayre2019replicable, sandve2013ten, chen2019open, lee_ten_2018, barba2019praxis} and are available online~\cite{arnold2019turing}. However, based on our results, we provide the following core recommendations for the researchers who use  R:

\begin{enumerate}
\item Capture library versions because R is not backward compatible. Consider using a configuration file such as \texttt{install.R}.
\item Use relative file paths as absolute (or full) file paths are a frequent cause of errors when re-executing code.
\item Simple bash files (or Make) could help automate your code and specify the execution sequence.
\item Consider using a research workflow (like Common Workflow Language~\cite{amstutz2016common}) to automate and parallelize your analysis. Many workflow managers also help in defining the runtime environment.
\item Test your code in a clean environment before sharing or publishing it, as it could help you identify dependencies and missing files. 
\item Use free and open-source software whenever possible, as proprietary software can hinder transparency and reproducibility of your research.
\end{enumerate}

Data and software repositories aim to provide high-quality resources that can be leveraged for further research. The following are our recommendations for data repository operators and curators:

\begin{enumerate}
\item Capturing re-execution commands for each research dataset would be immensely helpful for reusers.  It would resolve the ambiguity of the file execution order (workflow) and showcase its input arguments. Support for metadata fields or files capturing such commands does not currently exist but could be incorporated into dataset metadata at the repository.
\item Data repository integration with reproducibility platforms (such as Code Ocean~\cite{clyburne2019computational, cheifet2021promoting}, 
Whole Tale~\cite{brinckman2019computing}, Jupyter Binder~\cite{ragan2018binder}) and Renku\footnote{https://datascience.ch/renku/} could help capture library dependencies and test the code before it is published. It could be implemented as a part of the research submission workflow~\cite{trisovic_advancing_2020}. These tools use container technology that has been deemed valuable for preserving code~\cite{chuah2019documenting}.
\item Creating a working group that would support various aspects of reproducible research, investigate state of the art, and improve the quality of shared code would be beneficial. At Dataverse, we have created a \textit{Software, Workflows, and Containers} working group, which gathers experts, identifies community-wide problems, prioritizes them, and implements solutions in the Dataverse software. 
\end{enumerate}

Finally, our results suggest that journal data policy strictness positively correlates with the observed code re-execution rate. Therefore, journals play a critical role in making scholarly communication successful, and they have the power to require that the underlying data and code accompany articles. Our recommendations for the journal editors:

\begin{enumerate}
\item Consider implementing a simple review of all deposited material if a code verification is infeasible for your journal.
\item Create a reproducibility checklist that include code best practices (as recommended above) and testing code re-execution in a clean environment to make the submission and review process more straightforward~\cite{the_software_sustainability_institute_2018_2159713, goeva2020toward, pineau2020improving}.
\item Consider recommending the use of certain libraries or tools that facilitate code automation and re-execution. 
\end{enumerate}

\section{Related work}

Claims about a reproducibility crisis attracted attention even from the popular media, and many studies on the quality and robustness of research results have been performed in the last decade~\cite{national2019reproducibility, baker_1500_2016}. Most  reproducibility studies were done manually, where researchers tried to reproduce previous work by following its documentation and occasionally contacting original authors. Given that most of the datasets in our study belong to the social sciences, we reference a few reproducibility studies in this domain that emphasize its computational component (i.e., use the same data and code). Chang and Li attempt to reproduce results from 67 papers published in 13 well-regarded economic journals using the deposited supplementary material~\cite{chang2015economics}. They successfully reproduced 33\% of the results without contacting the authors and 43\% with the authors' assistance. Some of the reasons for the reduced reproducibility rate are proprietary software and missing (or sensitive) data. Stodden and collaborators conduct a study reporting on both reproducibility rate and journal policy effectiveness~\cite{stodden_empirical_2018}. They look into 204 scientific papers published in the journal Science, which previously implemented a data sharing policy. The authors report being able to obtain resources from 44\% of the papers and reproduce 26\% of the findings. They conclude that while a policy represents an improvement, it does not suffice for reproducibility. These studies give strength to our analysis as the success rates are comparable. Furthermore, by examining multiple journals with various data policy strictness, we corroborate the finding that open data policy is an improvement but less effective than code review or verification in enabling code re-execution and reproducibility.

Studies that focus primarily on the R programming language have been reported. Konkol and collaborators conducted an online survey among geoscientists to learn of their experience in reproducing prior work~\cite{konkol2019computational}. In addition, they conducted a reproducibility study by collecting papers that included R code and attempting to execute it. Among the 146 survey participants, 7\% tried to reproduce previous results, and about a quarter of those have done that successfully. For the reproducibility part of the study, Konkol and collaborators use RStudio and a Docker image tailored to the geoscience domain. They report that two studies ran without any issues, 33 had resolvable issues, and two had issues that could not be resolved. For the 15 studies, they contacted the corresponding authors. In total, they encountered 173 issues in 39 papers. While we cannot directly compare the success rate due to the different approaches, we note that much of the reported issues overlap. In particular, issues like a wrong directory, deprecated function, missing library, missing data, and faulty calls that they report are also frequently seen in our study.

Large-scale studies have the strength to process hundreds of datasets in the same manner and examine common themes.   Pimentel and collaborators retrieved over 860,000 Jupyter notebooks from the Github code repository and analyzed their quality and reproducibility~\cite{pimentel2019large}. The study first attempted to prepare the notebooks' Python environment, which was successful for about 788,813 notebooks. Out of those, 9,982 notebooks exceeded a time limit, while 570,476 failed due to an error. A total of 208,323 of the notebooks finished their execution successfully (24.11\%). About 4\% re-executed with the same result, which was inferred by comparing it with the existing outputs in the notebook. This result is comparable to the re-execution rate of 27\% in our previous analysis of Python code from Harvard Dataverse repository~\cite{trisovic2021repository}. We also note that Pimentel and collaborators performed the study on diverse Jupyter notebooks, which often include prototype development and educational coding. Our study is solely based on research code in its final (published) version. The studies are not directly comparable due to the use of different programming languages. However, we achieve a comparable result of 25\% when re-executing code without code cleaning. Also, the fact that the most frequent errors relate to the libraries in both studies signals that both programming languages face similar problems in software sustainability and dependency capture.

\section{Conclusion}

This paper presented a study addressing two aspects of research code quality: programming literacy and re-execution in the R programming language. Our primary contributions are collected data, open-source code, presented analysis findings, and a set of recommendations for researchers, data repositories, and academic journals that aim to improve code quality and research reproducibility. 

In the first part of the study, we saw that most datasets with R code on the Harvard Dataverse repository are relatively small, amounting to less than 10 MB and containing less than 15 files. About a third of the datasets have code in other programming languages (often proprietary), which needs to be considered for reproducibility and reuse. We find evidence of both good and bad coding practices. The majority of the datasets contain some sort of documentation and comments in the R code. In addition, dataset files and user-defined variables in the code often have long names (on average, 17 and 10 characters, respectively) and are thus more descriptive for a secondary user. However, the result reporting in Rnw and RMD is rarely used. We find that most used R libraries enable data visualization and analysis, but libraries that support code testing, provenance capture, and automated workflows are currently not used.

In the second part of the study, we probed the R research code against three different R versions and a code cleaning algorithm to get a total re-execution success rate of about 40\% (1,472 out of 3,695). We observed that curated data collections, particularly journal collections within the Harvard Dataverse repository, have higher re-execution rates. Furthermore, journals with stricter data policy (code review and verification) have the highest re-execution rates. We do not observe a significant correlation in success rates in datasets and R versions released in the same year, likely due to the inherent properties of R library management.

Lastly, we formulate several recommendations based on our findings that could have a measurable impact on code re-execution. We already saw that a basic code cleaning algorithm presented in this study improved code re-execution by about 10\%. Further, we suggest that developing and encouraging the use of a small number of tools (either directly by authors or integrated into data repositories) would have a similarly significant impact. Employing these recommendations would further help researchers, repositories, and journals contribute to research transparency and reproducibility.

\appendix

\section{Technical implementation of code cleaning} \label{cleaning}

Our code cleaning implementation aims to solve some of the most common errors and ensure that used libraries are installed. All R files are scanned and modified if a common problem is detected. Our code cleaning approach is relatively simple to minimize the chance of "breaking the code".\footnote{We do not use standardized static analysis packages such as \texttt{goodpractice} or \texttt{lintr} because they do not modify the code automatically.} First, all non-ASCII R files are converted to ASCII to reduce the chance of syntax error caused by symbols from other operating systems.

Even though it is considered a good practice in R, setting a working directory (setwd) function often causes errors if the directory path cannot be found. This is one of the things that our code cleaning implementation targets. It detects the \texttt{setwd} command and replaces it with a new one that makes the current directory with the downloaded files, a working directory.

Other file path errors were solved with the use of the basename function. The basename function ignores the path and the path separators. For example, if a long file path is used in the \texttt{read.csv} function, it will be ignored, and only the name of the target file will be used. Such an approach works in this scenario because all files from the replication package are downloaded one by one and stored in the same directory. In the implementation, a code line like: \texttt{file.path("/Dropbox/my\_datafile.csv")} would therefore be replaced with: 

\begin{lstlisting}[language=R]
basename(file.path("/Dropbox/my_datafile.csv"))
\end{lstlisting}

In the R programming, all libraries can be installed directly from the script using the \texttt{install.packages} command. Therefore if we detect that a needed library was not present in the working environment, we should add this command to install it and avoid an error. We tested a few approaches to identify the used and pre-installed libraries. Ultimately, using a combination of the functions \texttt{require} and \texttt{install.packages} proved to be the best solution (i.e., only if the library cannot be loaded with \texttt{require} it is installed from the code). The benefit of using the "require" function is that it returns a logical value by default, or "true" if the package is loaded and "false" if it is not. Therefore, we could check if the package is present and only install it if it is not. Such an approach saved time and reduced the chances of errors caused by duplicated code. As an example from the implementation, a line \texttt{library(dplyr)} would be replaced with: 

\begin{lstlisting}[language=R]
if (!require("dplyr")) install.packages("dplyr")
\end{lstlisting}

The US Cran\footnote{http://cran.us.r-project.org} was set as default to avoid the Cran errors.

\section*{Code Availability and Data Records}

Data and code are available on the Harvard Dataverse repository at \texttt{10.7910/ DVN/UZLXSZ} under \texttt{CC0} licence, and on GitHub at https://github.com/atrisovic/ dataverse-r-study under MIT license.

\section*{Acknowledgements}

This work is partially funded by the Sloan Foundation (award number G‑2018‑ 11111). A. Trisovic is funded by the Alfred P. Sloan Foundation (grant number P-2020-13988). This work was supported by the AWS Research Credits.

A. T. thanks Christopher Chen, Steven Worthington, Leonid Andreev and  Åsmund S. Folkestad for helpful conversations. The authors thank Christopher Chen and Margo Seltzer for their contributions. Thank you to the Dataverse team at Harvard's IQSS for their help and support.

\section*{Author contributions}

A.T.: conceptualization, software, formal analysis, investigation, writing-- original draft preparation, methodology, validation, visualization, writing-- review and editing, funding acquisition, data curation

M.K.L.: conceptualization, writing--review and editing

T.P.: conceptualization, writing--review and editing

M.C.: conceptualization, supervision, writing--review and editing, project administration, funding acquisition

\section*{Competing interests}

The authors have no conflicts of interest to declare.

\bibliographystyle{naturemag}
\bibliography{refs}

\end{document}